\documentclass{article}

\newcommand {\offs}      {\ensuremath{O}}
\newcommand {\Eptcl}     {\ensuremath{E_{\mathrm {ptcl}}}}
\newcommand {\Edet}      {\ensuremath{E_{\mathrm {cal} }}}
\newcommand {\kb}      {\ensuremath{k_{\mathrm {bias}}}}
\newcommand {\EstPtcl}     {\ensuremath{\hat{E}_{\mathrm{ptcl}}\left(\Edet\right)}}
\newcommand {\EstDet}      {\ensuremath{\hat{E}_{\mathrm{det}}\left(\Eptcl\right)}}
\newcommand {\pt}         {\ensuremath{p_T}}
\newcommand {\un}      {\,\mbox}
\newcommand {\GeV}      {\un{GeV}}

\newcommand {\ttbar}    {\ensuremath{t\tbar}}
\newcommand {\tbar}     {\ensuremath{\bar t}}
\newcommand {\W}        {\ensuremath{W}}
\newcommand {\Wbos}     {\ensuremath{W}~boson}

\newcommand {\emiss}{/\!\!\!\!E} 
\newcommand {\met} {\ensuremath{\emiss_{\rm T}}}
\newcommand {\mtop}    {\ensuremath{m_t}}
\newcommand {\nb}     {{\ensuremath{non-b}}}
\newcommand {\DZ}         {D0} 
\newcommand {\CDF}        {CDF}

\newcommand {\Eg}       {{\rm E.g.}}
\newcommand {\lxy}       {\ensuremath{L_{xy}}}
\newcommand {\IP}       {\ensuremath{\mathrm{IP}}}
\newcommand {\nbtag}    {\ensuremath{N_b}}

 \newcommand {\etal}     {{\it et al.}}

\newcommand{\halfWid}	{0.48}

\usepackage{graphicx}  
\title{Tools for top physics at \DZ}
\author{A.~Harel on behalf of the \DZ\ collaboration.}
\date{Proceedings of talk given at the \\ International Workshop on Top Quark Physics May 22, 2008}
\begin{document}

\maketitle

\begin{abstract}
Top quark measurements rely on the jet energy calibration and often on
$b$-quark identification. We discuss these and other tools and how they apply
to top quark analyses at \DZ. In particular some of the nuances
that result from \DZ's data driven approach to these issues are presented.
\end{abstract}

\section{Jet energy calibration}

Jet detection at \DZ\ is based on three finely segmented
liquid-argon and uranium calorimeters, hosted in
a central barrel and two end caps, 
that provide nearly full solid-angle coverage~\cite{ref:d0det}.
The calorimeters offer a stable response with good energy
resolution. Their total depth is more than $7.2$ interaction lengths.
In Run II of the Fermilab Tevatron Collider, \DZ\ calorimeters
collect charge within a time window of $260\un{ns}$.
They are partially compensating, with an electromagnetic response 
that is (roughly) 1.2 to 1.9 times higher than the hadronic response.

The region between the barrel and the end caps contains 
scintillator-based detectors that supplement 
the coverage of the main calorimeters. Jet reconstruction
in this region is inferior due to the complicated
geometry of the detectors and the large amount of passive material
(e.g. cryostat walls), but these effects are easily
accommodated within the data-driven techniques used in \DZ,
for example, the jet energy calibration described in this section.

Calorimeter readouts are grouped into pseudo-projective towers 
focused on the nominal interaction point for reconstruction purposes.
The energies deposited in calorimeter towers are then clustered 
into jets using the Run II iterative seed-based cone-jet
algorithm including mid-points~\cite{ref:d0jets}\ with 
cone radius 
${\cal R} = \sqrt{\left(\delta y\right)^2 + \left(\delta\phi\right)^2}
= 0.5$ in rapidity $y$ and azimuthal angle $\phi$.

\begin{figure}
\centering
\includegraphics[width=\halfWid\linewidth]{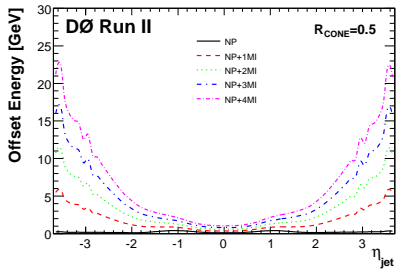}
\caption{\label{fig:offset}offset energy within the jet
for different primary vertex multiplicities, 
as a function of jet pseudorapidity.}
\end{figure}

The measured jet energies (\Edet) are calibrated to
match (on average) the energies at particle level (\Eptcl).
By ``particle level'' we refer to 
produced particles before they interact with material in the detector.
This calibration is usually described in terms of a multiplicative scaling
factor known as the jet energy scale ($\mathrm{JES}=\Eptcl / \Edet$)~\cite{ref:JES}.
\DZ\ parametrizes the calibrations for data and simulation (MC)
as
\begin{equation}
\Eptcl = \frac {\Edet - \offs}{R S} \cdot \kb,
\end{equation}
where the terms are as follows:
\begin{itemize}
\item the offset energy, \offs, is the energy 
not associated to the hard scatter: noise, pile-up, and
multiple collisions. Note that beam remnants and multiple parton interactions
(in the same collision) are not included.
It is calculated based on the energy density measured
in data from regions outside jets, as a function of the
number of primary vertices (PVs) reconstructed in the event
(see fig~\ref{fig:offset}).
\item the response, $R$, is the fraction of particle jet energy deposited in the calorimeter by the particles.
It is measured in three steps. First the photon energy scale
is calibrated with $Z\to e^+e^-$ data and with a detector simulation
 specifically tuned to reproduce electromagnetic showering,
which is used to translate between electron and photon energy scales.
Then the response in the central region is calibrated with photon+jet events
as a function of the jet energy.
Finally the response is extrapolated to other regions using dijet events 
with a central jet.
\item the detector showering correction, $S$, 
accounts for energy flow in and out of the calorimeter jet 
due to detector effects (finite calorimeter tower and hadron shower size, magnetic field).
Detector showering is estimated by fitting energy profile templates
to the data. The templates are derived from the simulation, one
describes energy in particles that belong to the jet, and the other 
the energy in particles that do not belong to the jet.
\item \kb\ represents corrections for biases of the method,
which are derived by comparing measured and desired values
of the \offs\ and $R$ terms in various MC samples.
\end{itemize}

\begin{figure}
\centering
\mbox{
  \includegraphics[width=\halfWid\linewidth]{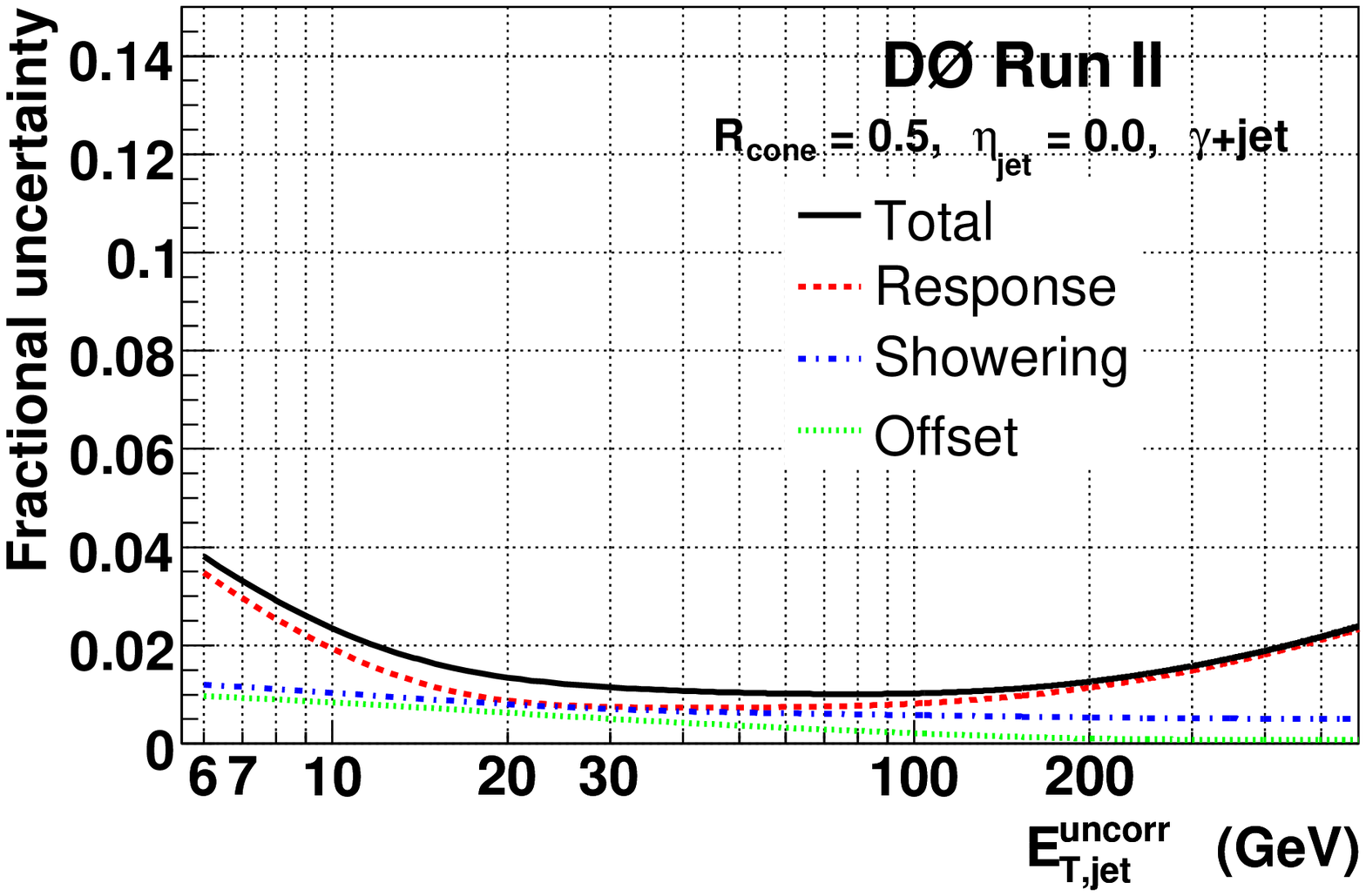}
  \includegraphics[width=\halfWid\linewidth]{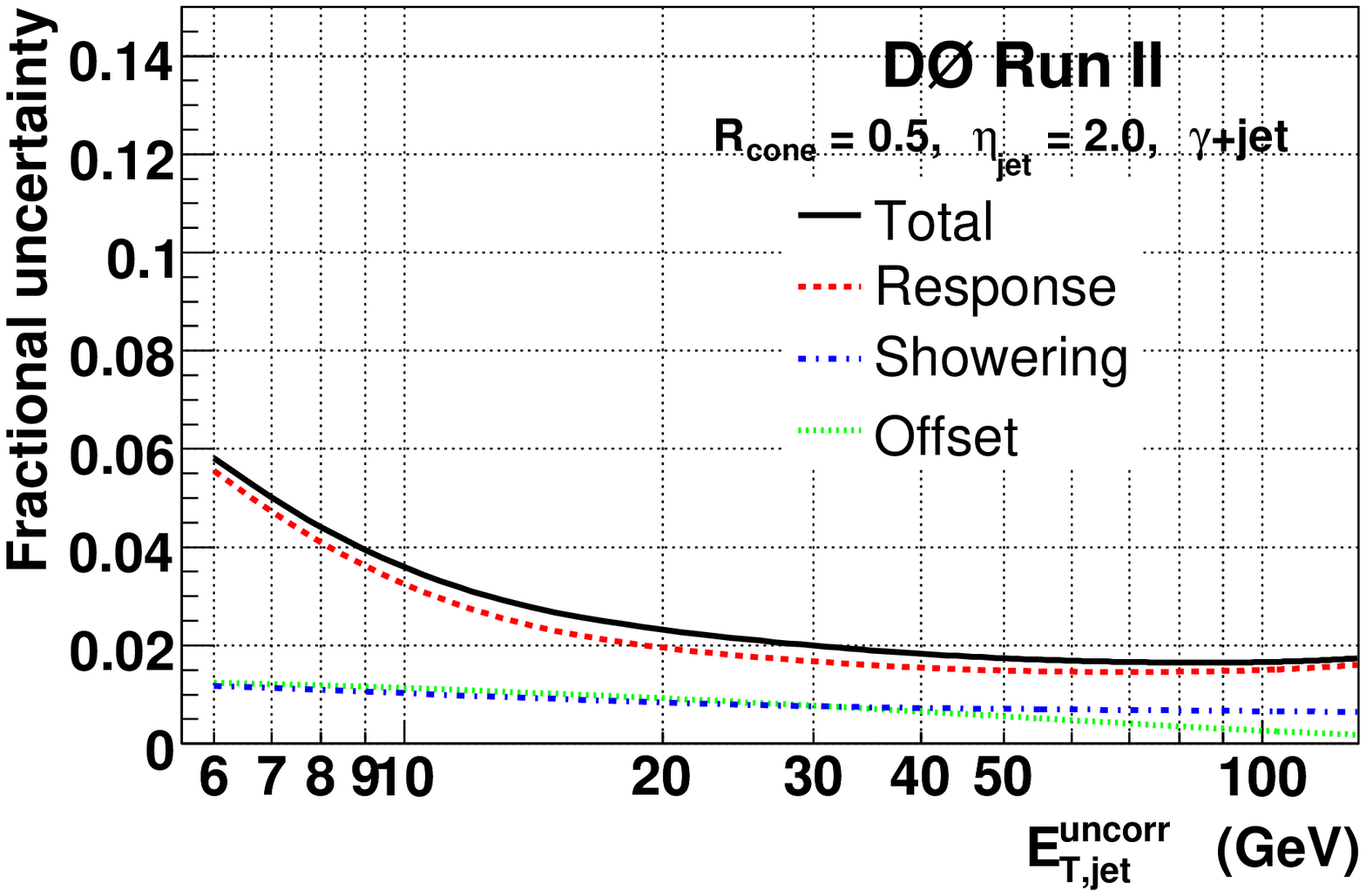}
}
\caption{\label{fig:jeserr}Fractional jet energy scale uncertainty for jets in data
         at eta=0.0 (left) and eta=2.0 (right), 
	 as a function of uncorrected jet transverse energy.}
\end{figure}

For central jets, the resulting JES decreases as 
a function of the jet energy from about 1.8 in data 
and 1.6 in simulation for $15\GeV$ jets,
to about 1.2 for the most energetic jets observed.
For forward jets the scales are a bit higher, but
the structure is similar. The resulting uncertainties
(shown in fig~\ref{fig:jeserr})
for central jets with transverse momenta (\pt) of 30-120\GeV\
are about 1\%. This unprecedented precision covers
most of the jet kinematics required for top quark measurements.
In that region, several components have comparable uncertainties;
in other regions the uncertainties on the response are bigger and
dominate the total uncertainty.

The jet energy scale benefits top analyses in several ways.
The most direct gain is that the JES puts jets collected 
at different regions of the detector
and with different instantaneous luminosities on an equal footing.
This improves the energy resolution and simplifies the analyses.
Another gain is that it puts jets from data and MC on an
equal footing. In fact, for almost all top analyses~\cite{ref:topasym}, 
it is only the relative (data over MC) JES that matters, rather than
the absolute JES. For example, top cross section analyses 
(e.g. \cite{ref:st} or~\cite{ref:Rb})
require the JES to calculate the signal selection efficiency, which
is taken from the MC. Similarly, top quark mass (\mtop) measurements~\cite{ref:mass}
use the MC's QCD modeling to translate the observed \Wbos\ 
and top quark mass peaks to their nominal (parton-level) masses,
and it is exactly in that translation that the relative JES is required.

The importance of the jet energy scale in top analyses can
be quantified by examining the impact of its uncertainties.
Though the JES is known to about 1\%, the resulting 
uncertainties in top cross section measurements are about
50\% of the total systematic uncertainties. For \mtop\ 
measurements they dominate the total systematic uncertainty.

\begin{figure}
\centering
\includegraphics[clip, bb=0 0 560 273, width=0.65\linewidth]{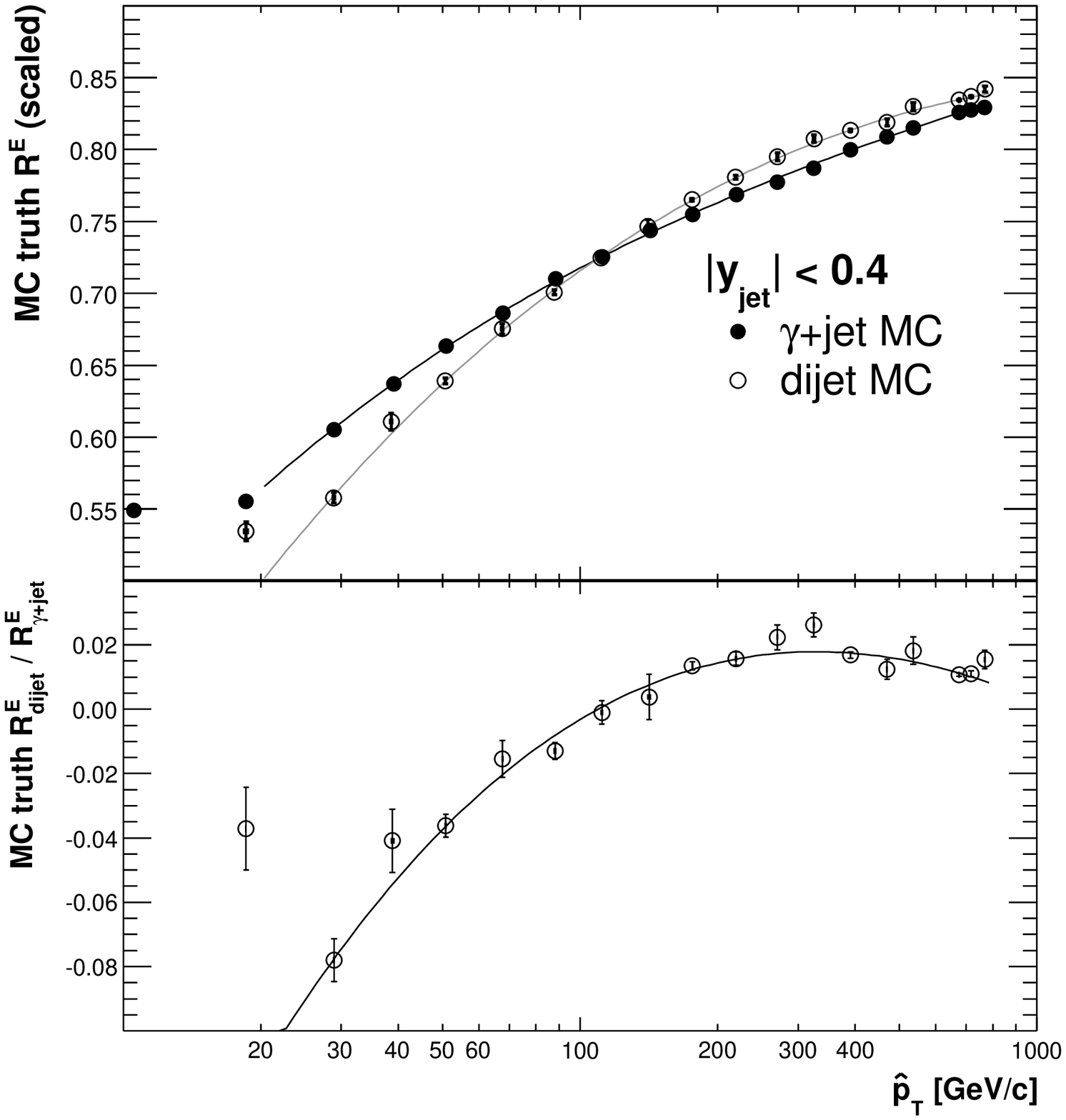}
\caption{\label{fig:dijet}
Estimated difference in response for jets from photon+jet
             and QCD dijet sample.}
\end{figure}

The final JES measurement, for the first $1\un{fb}^{-1}$
of \DZ\ data, 
is in some sense too precise: the uncertainty is so small that it 
is not directly applicable within its errors to jets from any but 
the photon plus jet sample. A detailed example, from
the inclusive jet cross section measurement~\cite{ref:mikko},
is the dijet energy scale. Since the hadronic response is particularly
low (relative to the electromagnetic response used to calibrate
the calorimeter) at low energies, and jets initiated by gluons have 
more particles and hence lower energies per particle than jets
initiated by quarks, the overall calorimeter 
response to gluon-initiated jets is about
5\% lower than the response to quark jets. At low \pt, the dijet sample is dominated
by gluon-initiated jets while the photon plus jet sample, on which the basic
JES was measured, is dominated by quark-initiated jets. This leads
to a $\approx5\%$ correction of the JES when applying it to
low \pt\ jets in a dijet sample (see fig~\ref{fig:dijet}), 
which is much larger than the uncertainty on the JES itself.

When applying the JES to top quark samples, additional complications appear.
The relative JES might differ between light and $b$ jets, due
to their different particle content (and spectra).
For our latest \mtop\ measurement this is the leading systematic
uncertainty since we fit both \mtop\ and an overall JES to the data.
The fitted JES is essentially from the light jets that make up
the observed \Wbos\ mass peak, but the fitted \mtop\ depends strongly
on the energy scale for the jet initiated by the $b$ quark from
the $t\to b \W$ decay.

Another complication is that the JES is defined as
\begin{equation}
\mathrm{JES}\left(\Edet\right) = \frac{\EstPtcl}{\Edet},
\end{equation}
where \EstPtcl\ is an unbiased estimator of
the corresponding particle-level jet energy.
This definition is rooted in QCD measurements, and
implicitly includes a bias appropriate to them: in a steeply falling spectrum
a symmetric finite resolution causes a bias as 
more events migrate 
into each \Edet\ bin from the heavily populated bin with slightly lower-\Edet\ 
than from the sparsely populated bin with slightly higher-\Edet.
The bias depends on the sample, and also on the resolution.
Thus the energy resolution and the sample (photon plus jet) are
implicit in the JES definition.
The former is particularly problematic for \DZ\ since 
the simulated jet resolutions are better than those observed in data.
The different resolutions are of course accounted for in analyses, 
but they also imply a different JES bias in data and MC
since the jet resolutions can only be measured (and calibrated)
after the JES is applied.

This raises the question:
Is this the best JES definition for top physics?
After all, the slope and resolution bias is almost irrelevant
for top samples due to the fairly flat jet energy spectra.
Why should we introduce this complication via the JES?
An alternative definition of the JES to be considered is:
\begin{equation}
\mathrm{JES}\left(\Eptcl\right) = \frac{\Eptcl}{\EstDet},
\end{equation}
where \EstDet\ is an unbiased estimator of
the corresponding detector-level jet energy.
Such a definition is independent of sample and energy
resolution, can easily be applied to data (using the
inverse function), and will improve the clarity of our
papers: currently a ``$20\GeV$'' jet in a \DZ\ top quark sample
corresponds on average to about $21\GeV$ at particle level
due to the sample dependent bias discussed above.

\section{Missing transverse energy likelihood}

The presence of non-interacting particles, such as neutrinos, in 
an event can be inferred from an imbalance in the transverse
components of the total momenta of the observable particles. 
In practice, we measure the imbalance observed in the 
calorimeter and refer to it as the missing transverse energy (MET).
Cuts on MET are used to enrich samples in top events with
a leptonically decaying \Wbos.
But due to the finite energy resolutions
multijet events can have
a sizable fake MET and are an important and difficult background
in many top analyses.

\begin{figure}
\begin{minipage}[b]{0.48\linewidth}
 \centering
 \includegraphics[clip, width=\linewidth]{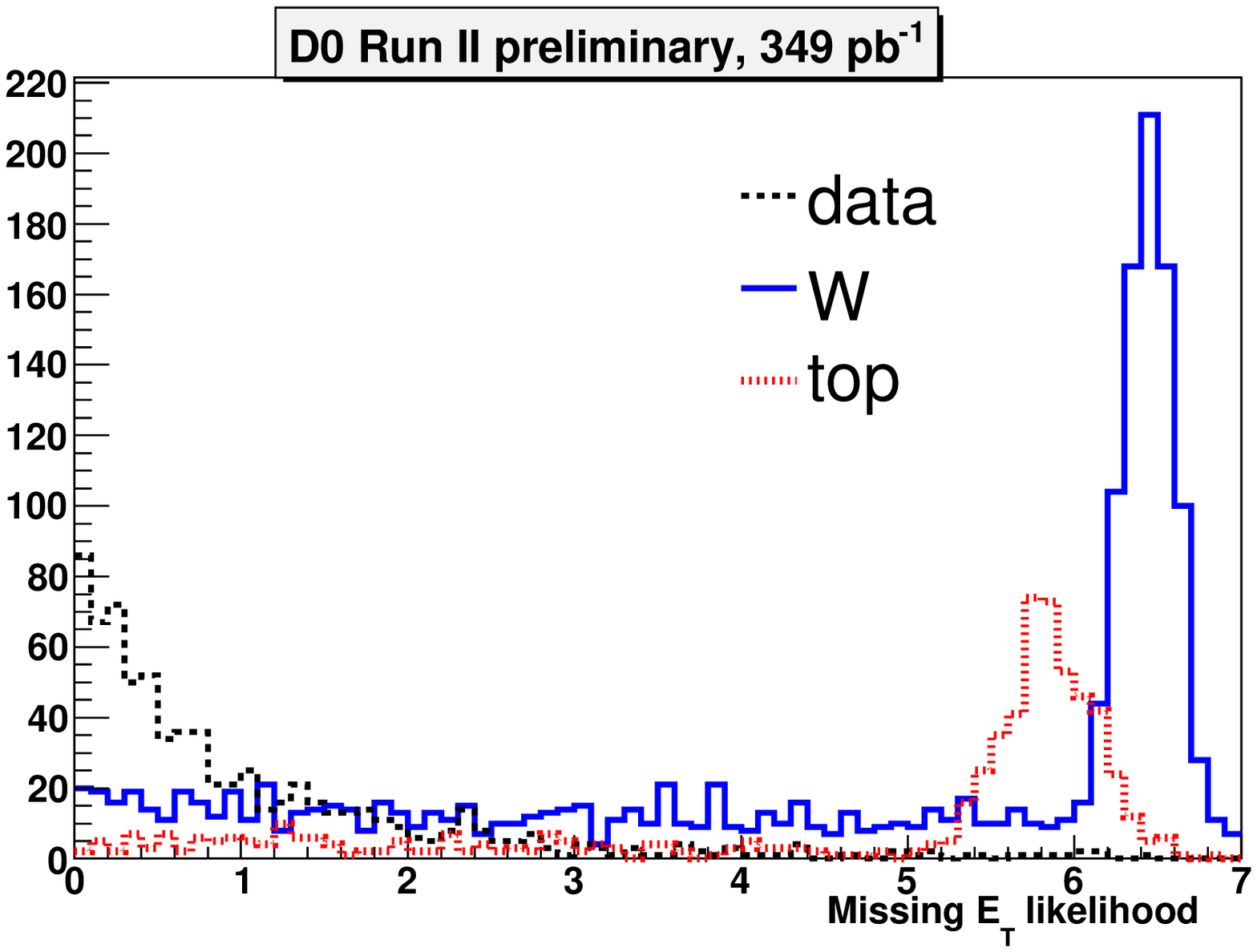}
 \caption{\label{fig:metl}
 \met\ likelihood for multijet dominated data (dashed black), \W\ plus jets (solid blue),
 and $\ttbar \to \tau \nu + \mathrm{jets}$ (finely dashed red).}
\end{minipage}
\hspace{0.04\linewidth}
\begin{minipage}[b]{0.48\linewidth}
 \centering
 \includegraphics[width=\linewidth]{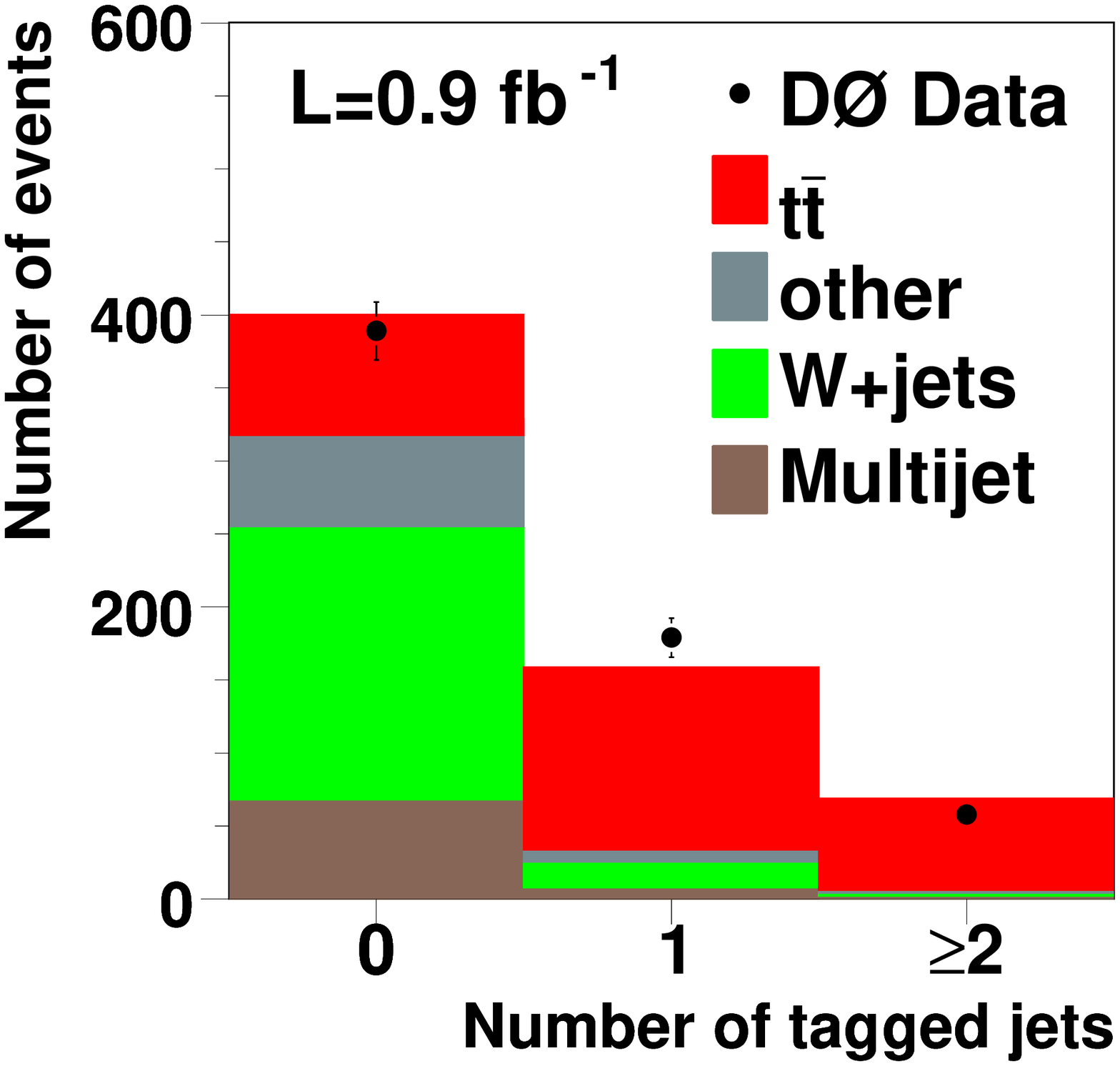}
 \caption{\label{fig:sample}
 Predicted and observed number of events 
 as a function of the number of $b$ tags, 
 for a typical top pair selection with 4 or more jets~\cite{ref:Rb}.}
\end{minipage}
\end{figure}

The MET-based background rejection is improved
by determining the MET resolution for each event
based on the detailed resolutions of the objects 
(jets, electrons and unclustered energy)
reconstructed in the event.
We then construct a log likelihood inspired
discriminating variable, that is the log of the probability
that the entire MET is a mismeasurement.
The construction also incorporates a ``soft'' limit
on high log likelihood values (see fig~\ref{fig:metl}).
This MET likelihood is a key tool in \DZ's top pair cross section
measurement in the $\tau$ plus jets channel~\cite{ref:taujets}.

\section{\boldmath $b$ tagging}

The identification of jets containing $b$ quarks is primarily
used in top analyses to suppress background, as the top signal contains $b$ jets, 
while most backgrounds do not. An example is shown
in fig~\ref{fig:sample}.
$b$ ID is also used to assign partons to jets,
often in conjunction with kinematic information. 
This is particularly needed in the 
fully-hadronic top pair decay channel where 90 
jet-parton assignments are possible for the canonical 6 jet events.
The average number of $c$ jets in these events is 1,
so that $c$-jet rejection is very useful when analyzing this channel.

The typical lifetime of a $B$ hadron is $1\un{ps}$, and
due to time dilation it can travel a few millimeters before
its decay. 
Here are the four basic $b$-tagging algorithms that are used in \DZ,
the first three are based on the $B$ hadron lifetime.
\begin{itemize}
\item The secondary vertex tagger (SVT) builds up track-based jets,
and for each one it selects tracks with high impact parameter
and attempts to build secondary vertices (SVs) from the selected tracks.
For each SV it calculates the decay length
significance $S\left(\lxy\right) = \lxy / \sigma\left(\lxy\right)$,
where \lxy\ is the visible decay length (the $z$ coordinate of the primary
vertex is known to a much lesser accuracy),
and $\sigma$ is its fitted
uncertainty. It then tags calorimeter jets matched within $\Delta R < 0.4$
to a track-jet with an SV with $S(\lxy)$ above some cut.
In what follows the cut is $S(\lxy)>3$.
\item The counting signed impact parameter tagger (CSIP) is based
on the tracks' impact parameter (IP) significance
$S\left(\IP\right) = \IP / \sigma\left(\IP\right)$.
$S(\IP)$ is a signed quantity, positive when the track's point of closest
approach to the PV is in the hemisphere defined around the 
track's reconstructed momentum with its origin at the PV. 
A jet is tagged as a $b$ jet 
if it has at least 2 tracks with $S(\IP) > 3$, or 
if it has at least 3 tracks with $S(\IP) > 2$.
\item The jet lifetime probability tagger (JLIP) translates each
track's $S(\IP)$ value into a probability that the track originated
at the PV, and then combines those probabilities into a jet-wide
probability.
\item The soft lepton tag (SLT) is based on $B$ hadron decay properties 
rather than on $B$ hadron lifetime. $B$ hadrons often decay into muons 
($B\left(b\to \mu X\right) \approx 11\%$), and since for reconstructed $b$ jets
$m_b << E_b$ the muons are usually collinear with the jet.
A jet is assigned an SLT if a muon is reconstructed within the jet. 
This tagger is very easy to model
and yields low systematic uncertainties that are completely different
than those that dominate the tracking based taggers. It also
identifies if the jet contained a $b$ quark or antiquark, which
can be useful~\cite{ref:wcjets}.
\end{itemize}

To combine the information from the three tracking tags,
we feed their outputs into a neural network (NN)
trained to discriminate between $b$ jets and light jets.
The dominant NN inputs are the decay length significance
of the SV, the weighted combination of the 
tracks' IP significance, and the JLIP output.
At a typical working point, the NN tagger tags $\approx 50\%$
of the $b$ jets, $\approx 10\%$ of the $c$ jets, and
$\approx 0.5\%$ of the light jets.

Since the performance of the tagging algorithm is difficult to simulate,
it is taken from data~\cite{ref:d0st}. It is split into two parts: taggability, which
is the probability that enough tracks were reconstructed (within $\Delta R = 0.5$
from the jet center) to $b$ tag the jet, and a tagging rate
(TR), that is an efficiency for $b$ and $c$ jets and a fake rate for light jets,
given that there are enough tracks reconstructed to $b$ tag the jet.

The taggability depends strongly on the $z$-coordinate of the PV,
as the PV may lie outside the fiducial volume of the \DZ\ silicon
tracker. Since all jets in an event are reconstructed as having
the same PV, this results in a large correlation between the jets
which must be described. 

The heavy flavor TRs given that the jet is taggable are
 measured using two base samples with different $b$ jet
contents: an inclusive sample of jets that contain a muon 
($n$) and a subsample of such jets that are back to back in
azimuth to a $b$-tagged jet ($p$).
Two almost uncorrelated $b$-tagging algorithms are used:
the track based algorithm under study and the SLT algorithm which
requires a muon within the jet.
The efficiencies are factorized into a function of \pt\ multiplied
by a function of $\eta$, 
and for each \pt\ or $\eta$ bin we have:
\begin{itemize}
\item eight event counts: $n$, $p$, $n^{NN}$,	$p^{NN}$, $n^\mu$, $p^\mu$,
$n^{NN \& \mu}$, and $p^{NN \& \mu}$, 
\item eight variables: $n_b$, $n_\nb$,
$p_b$, $p_\nb$, $\epsilon_b^\mu$, $\epsilon_\nb^\mu$, 
$\epsilon_b^{NN}$, and $\epsilon_\nb^{NN}$, where the $\epsilon$s
are efficiencies, and 
\item eight equations, such as $n=n_b + n_\nb$. The equations also
contain 4 corrections for possible correlations that are taken from the MC.
\end{itemize}
For each bin the solution of this equation system yields the TRs
for jets that contain muons. This is done for both data and MC, and the
MC is used to extrapolate the tag rates for all jets:
\begin{eqnarray}
\mathrm{TR}_b^{\mathrm{data}} & = & \frac {\mathrm{TR}_b^{\mathrm{MC}} \mathrm{TR}_b^{\mathrm{data}, \mu}}
                                          {\mathrm{TR}_b^{\mathrm{MC}, \mu}} \\
\mathrm{TR}_c^{\mathrm{data}} & = & \frac {\mathrm{TR}_c^{\mathrm{MC}} \mathrm{TR}_b^{\mathrm{data}, \mu}}
                                          {\mathrm{TR}_b^{\mathrm{MC}, \mu}}.
\end{eqnarray}
Similarly, fake rates ($\mathrm{TR}_l$) are measured in data using various negative tags,
for example, an SVT tag is considered as negative if the path from the PV to the
SV is in a direction opposite the momentum of the tracks in the SV.
MC corrections are then used to derive the fake rates from the negative tag rates.
The TRs are derived separately for each one of several working points,
for example, many top analysis use the NN output $> 0.65$ working point.

There are several strategies for using $b$-tagging information 
in top analyses. The standard strategy is to use the tagging rates
for a particular working point. 
If only the number of $b$ tags in each event (\nbtag) is of interest
the probability of having $\nbtag=0, 1, 2, \cdots$ is easily calculated
from the TRs (e.g. ref~\cite{ref:Rb}).
Several strategies are used when it is necessary to know which jets are $b$ tagged.
One can randomly assign a tag for each jet:
\begin{equation}
T_i = \left\{ \begin{array}{ll}
        \mathrm{true}  & \mbox{if $r \leq p_i$};\\
        \mathrm{false} & \mbox{if $r > p_i$},\end{array} \right.
\end{equation}
where $p_i$ is the jet's tagging probability and $r$ is a random
variable uniformly distributed between 0 and 1 (e.g. ref~\cite{ref:topasym}).
One can use all the possible assignments of $T_i$ values 
(for example ``$T_1$ $true$ and $T_j$ $false$ for $j>1$'' is a possible assignment),
giving each assignment a weight:
\begin{equation}
w_{T_1, T_2, \cdots} = \prod_{i} \left\{ \begin{array}{ll}
        p_i   & \mbox{if $T_i$ is $\mathrm{true}$};\\
        1-p_i & \mbox{if $T_i$ is $\mathrm{false}$}.\end{array} \right.
\end{equation}
(e.g. ref~\cite{ref:st}).
The latter method yields higher statistical strength, which unfortunately
is often hard to calculate due to the complicated correlations
between assignments derived from the same MC event.
Instead of using the TRs, it is also possible to weight the MC so
the tagging rates agree with data, which was done in ref~\cite{ref:mass}.

Another strategy that is currently being developed is to build
a semi-continuous $b$ tagger using rate functions  (for MC) for all working points.
The main difficulty to be resolved is how to account for systematic
correlations between the different bins.

An unusual strategy was used to measure the \Wbos\ helicity
in top decays~\cite{ref:Whel}:
since the analysis is sensitive only to the kinematic dependencies
of the TRs, and not to the overall rate, it was conceivable
that the known inaccuracies in the simulation of the tag rate
will not present a problem, as they have little kinematic dependence.
Thus, to utilize the full background rejection power of the NN tagger,
the highest NN outputs in the event were used as a discriminating variable. 
The difference between the simulated and actual distributions
was taken from a signal depleted sample and applied to the
selected sample to evaluate the resulting systematics.
The analysis proved to be quite insensitive to the missimulation,
and the resulting systematic uncertainty is only 10\% of the
total systematic uncertainties.

\section{Summary and outlook}

A successful top physics program requires well understood jet energy calibration and $b$ tagging.
In particular, the detector simulation might be a limiting factor. 
The unprecedented accuracy of \DZ's jet energy calibration 
raises sample dependence and detector simulation issues.
In this context it is interesting to compare \DZ's and \CDF's~\cite{ref:CDF}
``JES for top physics'' experience: perhaps a well calibrated 
parametrized MC is more useful than a full detector simulation?


\end{document}